\documentclass[prd,notitlepage,nofootinbib,twocolumn,preprintnumbers]{revtex4-1}
\usepackage{amsmath}
\usepackage{amssymb}
\usepackage{braket}
\usepackage{cancel}
\usepackage{color}
\usepackage[utf8]{inputenc}
\usepackage{slashed}
\usepackage[dvipsnames]{xcolor}
\usepackage[normalem]{ulem}
\usepackage{wasysym}
\usepackage[hidelinks]{hyperref}

\begin{document}

\newcommand{\order}[1]{\mathcal{O}\left(#1\right)}
\newcommand{\refapp}[1]{appendix~\ref{app:#1}}
\newcommand{\refeq}[1]{eq.~(\ref{eq:#1})}
\newcommand{\refsec}[1]{section~\ref{sec:#1}}
\newcommand{\reftab}[1]{table~\ref{tab:#1}}
\newcommand{\eps}{\varepsilon}

\title{A puzzle in $\boldsymbol{\bar{B}_{(s)}^0 \to D_{(s)}^{(*)+} \lbrace \pi^-, K^-\rbrace}$ decays and \\ extraction of the $\boldsymbol{f_s/f_d}$ fragmentation fraction}

\author{Marzia Bordone}
\email{marzia.bordone@uni-siegen.de}
\affiliation{Theoretische Physik 1, Naturwissenschaftlich-Technische Fakult\"at, Universit\"at Siegen, Walter-Flex-Stra{\ss}e 3, D-57068 Siegen, Germany}
\author{Nico Gubernari}
\email{nicogubernari@gmail.com}
\affiliation{Technische Universit\"at M\"unchen, James-Franck-Stra\ss{}e 1, 85748 Garching, Germany}
\author{Tobias Huber}
\email{huber@physik.uni-siegen.de}
\affiliation{Theoretische Physik 1, Naturwissenschaftlich-Technische Fakult\"at, Universit\"at Siegen, Walter-Flex-Stra{\ss}e 3, D-57068 Siegen, Germany}
\author{Martin Jung}
\email{martin.jung@unito.it}
\affiliation{Dipartimento di Fisica, Universit\`a di Torino \& INFN, Sezione di Torino, I-10125 Torino, Italy }
\author{Danny van Dyk}
\email{danny.van.dyk@gmail.com}
\affiliation{Technische Universit\"at M\"unchen, James-Franck-Stra\ss{}e 1, 85748 Garching, Germany}

\begin{abstract}
We provide updated predictions for the hadronic decays $\bar{B}_s^0\to D_s^{(*)+} \pi^-$ and $\bar{B}^0\to D^{(*)+} K^-$. They are based on $\mathcal{O}(\alpha_s^2)$ results for the QCD factorization amplitudes at leading power and on recent results for the $\bar{B}_{(s)} \to D_{(s)}^{(*)}$ form factors up to order ${\cal O}(\Lambda_{\rm QCD}^2/m_c^2)$ in the heavy-quark expansion. We give quantitative estimates of the matrix elements entering the hadronic decay amplitudes at order ${\cal O}(\Lambda_{\rm QCD}/m_b)$ for the first time. Our results are very precise, and uncover a substantial discrepancy between the theory predictions and the experimental measurements. We explore two possibilities for this discrepancy: non-factorizable contributions larger than predicted by the QCD factorization power counting, and contributions beyond the Standard Model. We determine the $f_s/f_d$ fragmentation fraction for the CDF, D0 and LHCb experiments for both scenarios.
\end{abstract}

\preprint{TUM-HEP 1271/20, P3H-20-034, SI-HEP-2020-17}

\maketitle

\section{Introduction}

The physics programs of the Large Hadron Collider experiments promise data sets of unprecedented sizes for a variety of $B_s$ decays. Consequently, analyses that emerge from these programs now dominate determinations of absolute branching fractions of $B_s$ decays. The biggest source of uncertainties in these analyses is the poorly known fraction of the $b$ quark fragmentation into $\bar{B}_s^0$ versus
$\bar{B}^0$ mesons, denoted as $f_s / f_d$. A promising approach to determine this ratio~\cite{Fleischer:2010ca} from data is the measurement of a ratio of branching fractions for hadronic $B$ decays:
\begin{align}
    \nonumber
        & \quad\, \frac{\sigma(pp \to \bar{B}_s^0 X) \times \mathcal{B}(\bar{B}_s^0 \to D_s^{(*)+} \pi^-)}{\sigma(pp \to \bar{B}^0 X) \times \mathcal{B}(\bar{B}^0 \to D^{(*)+} K^-)}\\
        & \equiv \frac{\sigma(pp \to \bar{B}_s^0 X)}{\sigma(pp \to \bar{B}^0 X)} \times \mathcal{R}^{P(V)}_{s/d}\,.
\end{align}
We consider the hadronic decays in this ratio very advantageous from the theory point of view. Since in these decays all valence quarks are distinguishable, we do not have to account for decay topologies involving penguin operators. For the same reason weak annihilation is not an issue either. 
Moreover, these decays are dominated by the color-allowed tree topology, and the color-suppressed operator enters only through perturbative or power corrections. In ref.~\cite{Huber:2016xod} 
the ratio that we here call $\mathcal R^P_{s/d}$ is given with a relative uncertainty of $\sim 9\%$.

The purpose of the present article is threefold. First, we revisit the QCD factorization framework for the non-leptonic decays $\bar{B}^0 \to D^{(*)+} K^-$ and $\bar{B}_s^0 \to D_s^{(*)+} \pi^-$ beyond leading power. We update the values of the $\bar{B}_q\to D_q^{(*)}$ form factors based on a recent analysis within the heavy-quark expansion (HQE)~\cite{Bordone:2019guc}. We also provide, for the first time, conservative numerical estimates for the necessary hadronic matrix elements that enter at next-to-leading-power in $\Lambda_{\rm QCD}/m_b$. Based on these improvements we predict the branching fractions of the four decays and, for the first time, use their theoretical correlation to reduce the uncertainty on their ratios $\mathcal R^{P(V)}_{s/d}$.

Second, we challenge existing experimental data on two-body non-leptonic decays into a heavy-light final state. We point out the subtleties in the comparison between theory and experiment.
Our analysis reveals a puzzling pattern in the comparison of theory predictions and data on the absolute branching fractions of the tree-dominated $\bar{B}_{(s)}^0 \to D_{(s)}^+ \lbrace \pi^-, K^-\rbrace$ decays,
whereas the predicted ratios $\mathcal{R}^{P(V)}_{s/d}$ are in good agreement with experiment.
We subsequently extract the $f_s / f_d$ fragmentation fraction for the CDF, D0 and LHCb experiments in a variety of scenarios.

Last but not least,
we critically assess possible origins for the observed puzzle --- which amounts to
a discrepancy of up to five standard deviations for the individual absolute branching fractions ---
with one of them being contributions from physics beyond the Standard Model (BSM).

This article is organised as follows. In section~\ref{sec:theory} we revisit the QCD factorization framework, including a thorough discussion of next-to-leading power hadronic matrix elements. We then discuss the numerical input parameters that enter our expressions, and give results for the non-leptonic branching fractions and the ratios $\mathcal{R}^{P(V)}_{s/d}$ at leading power. Our estimate of $\mathcal{R}^{P(V)}_{s/d}$ at next-to-leading power shows their robustness against power corrections. In section~\ref{sec:pheno} we compare the theoretical predictions to experimental data, explain in detail our extraction of the $f_s/f_d$  fragmentation fraction for various hadron colliders and uncover the puzzling pattern in non-leptonic $\bar{B}_{(s)}^0 \to D_{(s)}^+ \lbrace \pi^-, K^-\rbrace$ decays.
We determine possible solutions to this puzzle, among them effects from BSM physics.
We finally discuss prospects for future extractions of $f_s/f_d$ using the method discussed here and potential alternatives.
We conclude in section~\ref{sec:conclusion}. The article is supplemented by two appendices. In appendix~\ref{app:soft-gluon} we present the light-cone sum rule calculation for the soft-gluon matrix element relegated from section~\ref{sec:theory}, while in appendix~\ref{app:experimentalinputs} we give details on the experimental inputs.

\section{Theory prediction of \\ $\boldsymbol{\mathcal{B}(\bar{B}^0\to D^{(*)+}K^-)}$ and $\boldsymbol{\mathcal{B}(\bar{B}_s^0\to D_s^{(*)+} \pi^-)}$}
\label{sec:theory}

We begin by briefly summarizing the framework of collinear factorization (also known as QCD factorization (QCDF))  \cite{Beneke:2000ry}
for the decays $\bar{B}^0\to D^{(*)+} K^-$ and $\bar{B}_s^0 \to D_s^{(*)+} \pi^-$ in \refsec{theory:framework}.
Numerical results are presented subsequently in \refsec{theory:results}.

\subsection{Framework}
\label{sec:theory:framework}

Theory predictions for the decays under consideration are both relatively simple and particularly clean~\cite{Beneke:2000ry}. They are relatively simple, since neither penguin nor annihilation topologies contribute; they are particularly clean, again due to the absence of pollution from
weak annihilation, but also because no chirally enhanced hard-scattering contributions are present at order $\Lambda_\text{QCD}/m_b$. Since the latter two contributions constitute the main limitation of the QCDF approach, the resulting theory predictions are among the most reliable for non-leptonic decays. This is the basis for the phenomenological application of extracting the ratio $f_s/f_d$ from these modes \cite{Fleischer:2010ca}.
We emphasize that this statement does not hold for modes where the flavour of the spectator anti-quark is present in the valence content of the light meson. An example is the decay $\bar B^0\to D^+\pi^-$, which factorizes to leading power, but suffers from larger uncertainties due to endpoint divergences at subleading power.\\

The effective Lagrangian needed for the description of $\bar B^0\to D^{(*)+}K^-$ and $\bar B_s^0\to D_s^{(*)+}\pi^-$ decays reads
\begin{equation}
    \mathcal{L} = -\frac{4 G_F}{\sqrt{2}} V_{cb}^{\phantom{*}} V_{uq_2}^* \left[C_1 \mathcal{Q}_1^{q_2} + C_2 \mathcal{Q}_2^{q_2}\right]\,,
\end{equation}
where $q_2 = d, s$, and $V_{cb}$, $V_{uq_2}^*$ are CKM matrix elements. The Wilson coefficients $C_{1,2}$ are $q_2$-flavour universal in the Standard Model (SM). We choose these two current-current operators in the
CMM basis~\cite{Chetyrkin:1997gb}:
\begin{equation}
    \mathcal{Q}_{2(1)}^{q_2} \equiv \left[\bar{c} \gamma^\mu P_L (T^A) b\right]\, \left[\bar{q}_2 \gamma_\mu P_L (T^A) u\right]\,.
\end{equation}
The Wilson coefficients $C_{1,2}$ are known to next-to-next-to-leading logarithmic accuracy \cite{Gorbahn:2004my}, and at the
scale $\mu = m_b$ we use 
\begin{align}
    C_2 & = +1.010\, &
    C_1 & = -0.291\,.
\end{align}
Their uncertainties are negligible compared to those of the hadronic matrix elements.\\

In this work, the light pseudoscalar ($\bar u q_2)$ bound state is denoted as $L$ when the particular quark flavour $q_2$ is not relevant to the discussion.
For the discussion below we adopt the power-counting $\eps \sim \Lambda_\text{QCD} / E_L \sim \Lambda_\text{QCD} / m_b$.
Following this power counting the matrix elements of the current-current operators  factorize to leading power~\cite{Beneke:2000ry}:
\begin{align}
    \bra{D_q^{(*)+} L^-} \mathcal{Q}_i \ket{\bar{B}_q^0}
        = & \sum_{j} F_j^{\bar{B}_q \to D_q^{(*)}}(M_L^2) \nonumber\\
        \times&\int_0^1 du\, T_{ij}(u) \phi_L(u)+ \order{\frac{\Lambda_\text{QCD}}{m_b}}\,.
\end{align}
The amplitudes to leading power in $\eps$ read
\begin{equation}
\label{eq:amp:leading-power}
\begin{aligned}
    \mathcal{A}(\bar{B}_{q}^0 \to D_{q}^+ L^-)
        & = i \frac{G_F}{\sqrt{2}} V_{uq_2}^* V_{cb}^{\phantom{*}} a_1(D_q^+ L^-) f_L \\
        & \times F_0^{\bar{B}_q \to D_q}(M_L^2) (M_{B_q}^2 - M_{D_q}^2)\,,\\
    \mathcal{A}(\bar{B}_{q}^0 \to D_{q}^{*+} L^-)
        & = -i \frac{G_F}{\sqrt{2}} V_{uq_2}^* V_{cb}^{\phantom{*}} a_1(D_q^{*+} L^-) f_L \\
        & \times A_0^{\bar{B}_q \to D_q^*}(M_L^2) 2 M_{D_q^*} \eps^*(\lambda=0)\cdot q
\end{aligned}
\end{equation}
for either a pseudoscalar $D_q$ meson, or a longitudinal ($\lambda=0$) $D_q^*$ vector meson.
The structure of these amplitudes holds to all orders in $\alpha_s$. The effective Wilson coefficients
$a_1(D_q^{(*)+} L^-)$ have been computed to next-to-next-to-leading order in $\alpha_s$ in ref.~\cite{Huber:2016xod}.
We emphasize that the $SU(3)_F$ breaking of the amplitudes is numerically driven by the decay
constants of the light-meson. The deviation from the symmetry limit in heavy-to-heavy form factors
is additionally suppressed by the heavy-quark masses and therefore expected to be very small~\cite{Jenkins:1992qv,Boyd:1995pq,Kobach:2019kfb}.
A recent analysis of the symmetry breaking in heavy-to-heavy form factors yields results that are compatible
with this expectation~\cite{Bordone:2019guc}. To ascertain the stability of QCDF predictions
for $\mathcal R_{s/d}$, control of the symmetry-breaking effects is essential, even if they only arise within
power corrections.\\

The formulas for the $SU(3)_F$ ratios with either a pseudoscalar (P) or  vector (V) meson $D_q^{(*)}$ read
\begin{align}
  \mathcal  R^P_{s/d} & = \frac{\mathcal{B}(\bar{B}_s^0 \to D_s^{+} \pi^-)}{\mathcal{B}(\bar{B}^0 \to D^{+} K^-)} \nonumber\\
        & = \frac{\tau_{B_s}}{\tau_{B_d}}\left|\frac{V_{ud}}{V_{us}}\right|^2 
        \frac{f_\pi^2}{f_K^2}
         \left|\frac{F_0^{\bar{B}_s \to D_s}(M_\pi^2)}{F_0^{\bar{B} \to D}(M_K^2)}\right|^2
           \left|\frac{a_1(D_s^+ \pi^-)}{a_1(D^+ K^-)}\right|^2\nonumber\\
        & \times \left(\frac{M_{B_s}^2 - M_{D_s}^2}{M_B^2 - M_D^2}\right)^2 \frac{M_B^3}{M_{B_s}^3} \frac{\sqrt{\lambda(M_{B_s}^2, M_{D_s}^2, M_\pi^2)}}{\sqrt{\lambda(M_B^2, M_D^2, M_K^2)}}\,,
        \label{eq::RPsd}\\
    \mathcal R^V_{s/d} & = \frac{\mathcal{B}(\bar{B}_s^0 \to D_s^{*+} \pi^-)}{\mathcal{B}(\bar{B}^0 \to D^{*+} K^-)} \nonumber\\
        & = \frac{\tau_{B_s}}{\tau_{B_d}}\left|\frac{V_{ud}}{V_{us}}\right|^2 
        \frac{f_\pi^2}{f_K^2}
         \left|\frac{A_0^{\bar{B}_s \to D_s^*}(M_\pi^2)}{A_0^{\bar{B} \to D^*}(M_K^2)}\right|^2
           \left|\frac{a_1(D_s^{*+} \pi^-)}{a_1(D^{*+} K^-)}\right|^2\nonumber\\
        & \times \frac{M_B^3}{M_{B_s}^3} \frac{[\lambda(M_{B_s}^2, M_{D_s^*}^2, M_\pi^2)]^{3/2}}{[\lambda(M_B^2, M_{D^*}^2, M_K^2)]^{3/2}}\,.
        \label{eq::RVsd}
\end{align}
Here $\lambda(a,b,c) = a^2+b^2+c^2-2ab-2ac-2bc$ is the K\"all\'en function, and the formulas for the individual branching ratios can be found in~\cite{Beneke:2000ry}.
To test the consistency of the QCDF predictions, we also study the fixed-flavour ratios
of branching fractions to vector $D_q^{*+}$ mesons over pseudoscalar $D_q$ mesons:
\begin{align}
    \mathcal R_s^{V/P} &= \frac{\mathcal B(\bar B_s^0\to D_s^{*+}\pi^-)}{\mathcal B(\bar B_s^0\to D_s^{+}\pi^-)}\,,\\
    \mathcal R_d^{V/P} &= \frac{\mathcal B(\bar {B}^0\to D^{*+}K^-)}{\mathcal B(\bar B^0\to D^{+}K^-)}\,,
\end{align}
which have analogous expressions.\\

Power corrections to the amplitudes in \refeq{amp:leading-power} arise from a variety of effects. To order $\eps$,
these can potentially include higher-twist corrections to the light-meson light-cone distribution amplitude (LCDA); emission of a hard-collinear gluon from the spectator
$q$, or from the heavy bottom and charm quarks; and exchange of a soft gluon between the $\bar{B}_q^0\to D_q$ system
and the light meson. We briefly discuss each of these effects and provide updated numerical estimates for their magnitudes. Importantly, using form factors in terms of QCD quark fields in \refeq{amp:leading-power} avoids explicit $\Lambda/m_c$ power corrections. We note that implicit corrections due to application of the heavy-quark expansion to the form factors are part of our uncertainty budget as discussed below.\\

The contributions of the two-particle twist-three light-meson LCDA are suppressed with respect to the leading-twist contribution
by one power of $\eps$. As discussed in the literature, see e.g. ref.~\cite{Ball:2006wn}, twist-three corrections for
a pseudoscalar scale like $\mu_L / E_L$, which can be numerically large due to the normalization in terms of the factor
\begin{equation}
    \mu_L \equiv \frac{M_{L^-}^2}{m_u + m_{q_2}}\,.
\end{equation}
However, the twist-three corrections to the hadronic amplitude vanish algebraically at leading-order in $\alpha_s$~\cite{Beneke:2000pw}.
We find that higher-twist corrections only enter at order $\alpha_s\eps^2$ or higher.\\

Up to $\order{\eps}$, the exchange of a gluon between the heavy-to-heavy transition and the light meson can possibly enter in
one of three ways~\cite{Beneke:2000pw}.\\
$i)$~The exchange of a hard gluon from the $b$ or $c$ quark is a purely perturbative effect, and
accounted for by the $\order{\alpha_s}$ correction to the effective Wilson coefficients $a_1(D_q^{(*)} L)$.\\
$ii)$~The exchange of a hard-collinear gluon from the spectator quark is incompatible with the physical picture of the spectator having a soft momentum inside both the $\bar{B}_q^0$ and the $D_q$ meson.
The exchange of a hard-collinear gluon from the $b$ or $c$ quark is possible, and contributes through
a quark-antiquark-gluon Fock state of the light meson through three-particle LCDAs. Due to the $V-A$ structure
of the weak interaction the twist-three three-particle contribution is absent, and
the first contribution emerges at the twist-four level~\cite{Beneke:2000ry}. Within our power counting, the
twist-four contribution enters at the $\order{\eps^2}$ level. This power correction breaks $SU(3)_F$ symmetry
maximally, since it is absent for the $L=\pi^-$ case by virtue of G parity, but has a finite contribution
for the $L=K^-$ case. Using the expression for these corrections in ref.~\cite{Beneke:2000ry}, a model based on the
conformal expansion of the LCDAs and the numerical results of ref.~\cite{Ball:2006wn} we obtain
\begin{equation}
    \label{eq::NLP1}
    \frac{\mathcal{A}(\bar{B}^0 \to D^{(*)+} K^-)\big|_\text{NNLP(hc)}}{\mathcal{A}(\bar{B}^0 \to D^{(*)+} K^-)\big|_\text{LP}}
        \simeq \frac{C_1}{a_1} \times -0.64\%\,.
\end{equation}

  This term has been computed in the limit $m_c \ll m_b$. Analysis of
  eq. (56) of ref.~\cite{Beneke:2000ry} reveals that corrections due to a non-zero charm mass scale like $\order{\frac{\Lambda_\text{had}^2m_c}{m_b^3}}$,
  which are negligible to the accuracy for which we aim.\\
$iii)$~The exchange of a single soft gluon between the heavy-to-heavy transition and the light meson causes one of the
light quarks in the light meson to become hard-collinear~\cite{Beneke:2000ry}. This configuration can be
expressed through the heavy-to-heavy matrix elements of a non-local $\bar{c} G b$ current. We estimate the relevant matrix
elements using Light-Cone Sum Rules (LCSRs) in \refapp{soft-gluon}. Using these estimates, we obtain
ranges for the $\order{\eps}$ soft-gluon correction:
\begin{align}
    \frac{\mathcal{A}(\bar{B}_q^0 \to D_q^+ L^-)\big|_\text{NLP}}{\mathcal{A}(\bar{B}_q^0 \to D_q^+ L^-)\big|_\text{LP}}
        & \simeq \frac{C_1}{4a_1} \times [0.76, 7.6]\%\,,\\
    \frac{\mathcal{A}(\bar{B}_q^0 \to D_q^{*+} L^-)\big|_\text{NLP}}{\mathcal{A}(\bar{B}_q^0 \to D_q^{*+} L^-)\big|_\text{LP}}
        & \simeq \frac{C_1}{4a_1} \times [0.46, 4.6]\%\,.\label{eq::NLP2}
\end{align}
The lower value of these ranges corresponds to the sum rule results for our nominal inputs as discussed in \refapp{soft-gluon}.
To obtain a more conservative estimate, we increase the values of the scale-setting parameters $\lambda_E^2$ and $\lambda_H^2$ by one order of magnitude,
yielding the higher of the values.
We emphasize that the ratio of the hadronic matrix elements are matching the size expected from naive dimensional analysis.
However, the prefactor of $C_1 / a_1 \sim -1/3$ renders all estimates for the power corrections small, and supports the picture of these decays being very clean.

\subsection{Numerical Inputs and Results}
\label{sec:theory:results}

\noindent
\textbf{Form factors}\quad
For the form factors and form-factor ratios entering our predictions, we use
the posterior samples for the HQE parametrization from ref.~\cite{Bordone:2019guc}. This parametrization includes contributions to orders $1/m_b,1/m_c^2,\alpha_s$, thereby providing a consistent treatment of the form factors and specifically also their uncertainties including $1/m_c^2$ contributions, see refs.~\cite{Bordone:2019guc,Bordone:2019vic} for details.
In \reftab{theory-inputs+results}~the values of the form factors used in this work are collected together with the ones used in ref.~\cite{Huber:2016xod}. We notice that with respect to ref.~\cite{Huber:2016xod} these form factor predictions have reduced uncertainties, especially for the $\bar{B}_q^0\to D^+_q$ cases. This is dominantly due to improved lattice QCD determinations of the corresponding form factors presented in refs.~ \cite{McLean:2019qcx,Lattice:2015rga}, which are the most constraining inputs for the analysis in ref.~\cite{Bordone:2019guc}. 
The correlation coefficients between these form factor values are
\begin{equation}
\newcommand{\pp}{\phantom{+}}
\begin{pmatrix}
    \pp 1.0000 & \pp 0.0814 & \pp 0.1702 & -   0.0255 \\
    \pp 0.0814 & \pp 1.0000 & \pp 0.0059 & \pp 0.4787 \\
    \pp 0.1702 & \pp 0.0059 & \pp 1.0000 & \pp 0.0134 \\
    -   0.0255 & \pp 0.4787 & \pp 0.0134 & \pp 1.0000   
\end{pmatrix}\,,
\end{equation}
with rows and columns ordered as $F_0^{\bar{B} \to D}(M_K^2)$, $A_0^{\bar{B} \to D^*}(M_K^2)$, $F_0^{\bar{B}_s \to D_s}(M_\pi^2)$, and $A_0^{\bar{B}_s \to D_s^*}(M_\pi^2)$.
Accounting for these correlations, we obtain the form factor ratios:
\begin{align}
    \label{eq:ratio-F0sF0d}
    \left|\frac{F_0^{\bar{B}_s \to D_s}(M_\pi^2)}{F_0^{\bar{B} \to D}(M_K^2)}\right|
        & = 1.001 \pm 0.021\,,\\
    \label{eq:ratio-A0sA0d}
    \left|\frac{A_0^{\bar{B}_s \to D_s^*}(M_\pi^2)}{A_0^{\bar{B} \to D^*}(M_K^2)}\right|
        & = 0.9729 \pm 0.080\,,\\
    \label{eq:ratio-F0sA0s}
    \left|\frac{F_0^{\bar{B}_s \to D_s}(M_\pi^2)}{A_0^{\bar{B}_s \to D_s^*}(M_\pi^2)}\right|
        & = 0.9773 \pm 0.095\,,\\
    \label{eq:ratio-F0dA0d}
    \left|\frac{F_0^{\bar{B}   \to D}(M_K^2)}{A_0^{\bar{B} \to D^*}(M_K^2)}\right|
        & = 0.9507 \pm 0.095\,.
\end{align}

\noindent
\textbf{Effective Wilson coefficients}\quad
Based on the analytic results for the effective Wilson coefficients to NNLO in $\alpha_s$ from ref.~\cite{Huber:2016xod},
we produce numbers for their absolute values. The numbers used in the present work contain more significant digits, but are otherwise identical to those in ref.~\cite{Huber:2016xod}, see \reftab{theory-inputs+results}.
In addition, we require the following three ratios of effective Wilson coefficients:
\begin{align}
    \left|\frac{a_1(D_s^+ \pi^-)}{a_1(D^+ K^-)}\right|
        & = 1.0024^{+0.0023}_{-0.0011}\, , \\[0.8em]
    \left|\frac{a_1(D_s^+ \pi^-)}{a_1(D_s^{\ast+} \pi^-)}\right|
        & = 1.0013^{+0.0006}_{-0.0005}\, , \\[0.8em]
    \left|\frac{a_1(D^+ K^-)}{a_1(D^{\ast+} K^-)}\right|
        & = 1.0013^{+0.0005}_{-0.0005}\, .
\end{align}

\noindent
$\boldsymbol{|V_{cb}|}$\quad For $|V_{cb}|$, we use the value obtained in refs.~\cite{Bordone:2019guc,Bordone:2019vic},
which constitutes an average of the values extracted from inclusive and exclusive semileptonic decays.
This value is larger than in ref.~\cite{Huber:2016xod} where the value for $|V_{cb}|$ from exclusive decays as of 2016 was used.\\

\noindent
\textbf{$\boldsymbol{|V_{uq_2}|}$ and light-meson decay constants}\quad
Throughout this work, we use directly the products of the CKM matrix elements $|V_{ud}|$ or $|V_{us}|$ with
the respective light-meson decay constants $f_\pi$ or $f_K$ as obtained by the PDG from leptonic $\pi^-,K^-$ decays~\cite{Tanabashi:2018oca}. Our current values together with those from ref.~\cite{Huber:2016xod} are again collected in \reftab{theory-inputs+results}. 
Due to the partial cancellation of radiative corrections their ratio is even more precisely known than the individual quantities,
\begin{equation}
    \left|\frac{V_{ud}}{V_{us}}\right|^2 \times \left|\frac{f_\pi}{f_K}\right|^2
        = 13.128 \pm 0.038\,,
\end{equation}
leaving a negligible relative uncertainty of $1.5\permil$.\\

\noindent
\textbf{Leading-power predictions and comparison}\quad
Using the above inputs we obtain for the QCDF results at leading power (LP) in $\eps$
the values in \reftab{theory-inputs+results}, again listed together with the results
of ref.~\cite{Huber:2016xod}. The most significant changes arise due to the shifts in $|V_{cb}|$ ($+4\%$), and the $\bar{B}_s^0\to D_s^{(*)}$ form factors. The former shift is mostly due to our inclusion of the $|V_{cb}|$ value obtained from inclusive decays \cite{Gambino:2016jkc},
while the latter ($-4\%$ for $F_0$ and $+33\%$ for $A_0$) is caused by improved calculations of the $\bar{B}_s^0\to D_s^{(*)}$ form factors \cite{McLean:2019qcx,McLean:2019sds,Gubernari:2018wyi} which have become available since the analysis of ref.~\cite{Huber:2016xod}, where the only available calculation~\cite{blasi:1993fi} as of 2016 was used. The resulting shifts in the absolute branching fractions follow correspondingly,
with $+8\%$ for the $\bar{B}_d^0$ branching fractions, no significant shift in $\bar{B}_s^0\to D_s^+ \pi^-$,
and $+90\%$ in $\bar{B}_s^0\to D_s^{*+} \pi^-$. Note, that the current form factor results shift the QCDF predictions from $2.24^{+0.56}_{-0.50}$ (in units of $10^{-3}$) to the new value $4.30^{+0.9}_{-0.8}$
which is now $2.3\sigma$ away from the experimental value $2.1 \pm 0.5$~from the third column of table~\ref{tab:resultsBR}.
The uncertainties of all our predictions are dominated by the form factor uncertainties, which will be reduced in the future. In the case of the absolute branching fractions of modes with $D_q$ final states, the uncertainties from $|V_{cb}|^2\sim 2.4\%$ and $|a_1|^2\sim 2.4\%$ are comparable to the form factor ones ($\sim 3.3\%)$, while for absolute branching fractions for $D_q^*$ final states and ratios of branching fractions the form factor uncertainties dominate by far.

In addition we obtain, still to leading power,
the following ratios of branching fractions:
\begin{align}
    \mathcal R_{s/d}^P\big|_\text{LP}
        & = 13.5^{+0.6}_{-0.5}\,,\\
    \mathcal R_{s/d}^V\big|_\text{LP}
        & = 13.1^{+2.3}_{-2.0}\,,\\ 
    \mathcal R_{s}^{V/P}\big|_\text{LP}
        & = 0.97^{+0.20}_{-0.17}\,,\\ 
    \mathcal R_{d}^{V/P}\big|_\text{LP}
        & = 1.01\pm0.11\,.
\end{align}
Two of these values can be compared to the ones in ref.~\cite{Huber:2016xod} where
\begin{align}
    \mathcal R_{s/d}^P\big|_\text{LP}
        & = 14.67^{+1.34}_{-1.28}\,,\\
    \mathcal R_{d}^{V/P}\big|_\text{LP}
        & = 0.863^{+0.158}_{-0.147}
\end{align}
are obtained. We observe that the central values are compatible within uncertainties, and that in our new analysis the latter are $55\%$ and $30\%$ smaller, respectively.\\

\noindent
\textbf{Next-to-leading power predictions}\quad
When including the hadronic matrix elements at next-to-leading power (NLP) in $\Lambda_\text{QCD}/m_b$,
we obtain:
\begin{align}
    \mathcal R_{s/d}^P\big|_\text{NLP}/\mathcal R_{s/d}^P\big|_\text{LP}-1
        & \approx -1.7\permil\,,\\
    \mathcal R_{s/d}^V\big|_\text{NLP}/\mathcal R_{s/d}^V\big|_\text{LP}-1       & \approx -1.7\permil\,,\\ 
    \mathcal R_{s}^{V/P}\big|_\text{NLP}/\mathcal R_{s}^{V/P}\big|_\text{LP}-1
        & = (0.4\ldots 8.2)\permil\,,\\ 
    \mathcal R_{d}^{V/P}\big|_\text{NLP}/\mathcal R_{d}^{V/P}\big|_\text{LP}-1
        & = (0.4\ldots 8.2)\permil\,.
\end{align}
The shift from LP to NLP is negligible, as anticipated in \refsec{theory:framework}.
This underscores why the QCDF predictions for these decays are considered to be among
the most reliable for two-body $B$ decays.

\begin{table}
\begin{center}
\renewcommand{\arraystretch}{1.3}
    \begin{tabular}{ c c c c }
    \toprule
         quantity                                   & unit         & this work                        &ref.~\cite{Huber:2016xod} ($2016$) \\
    \hline
         $F_0^{\bar B\to D}(M_K^2)$                 & ---          & $0.672\pm 0.011$              & $0.670 \pm 0.031$ \\
         $F_0^{\bar B_s^0\to D_s}(M_\pi^2)$           & ---          & $0.673\pm 0.011$              & $0.700 \pm 0.100$ \\
         $A_0^{\bar B\to D^*}(M_K^2)$               & ---          & $0.708\pm 0.038$              & $0.654 \pm 0.068$  \\
         $A_0^{\bar B_s^0\to D_s^*}(M_\pi^2)$         & ---          & $0.689\pm 0.064$              & $0.520 \pm 0.060$  \\
    \hline
         $\left|a_1(D_s^+ \pi^-)\right|$            & ---          & $1.0727^{+0.0125}_{-0.0140}$  & $1.073^{+0.012}_{-0.014}$ \\
         $\left|a_1(D^+ K^-)\right|$                & ---          & $1.0702^{+0.0101}_{-0.0128}$  & $1.070^{+0.010}_{-0.013}$ \\
         $\left|a_1(D_s^{*+} \pi^-)\right|$         & ---          & $1.0713^{+0.0128}_{-0.0137}$  & $1.071^{+0.013}_{-0.014}$ \\
         $\left|a_1(D^{*+} K^-)\right|$             & ---          & $1.0687^{+0.0103}_{-0.0125}$  & $1.069^{+0.010}_{-0.013}$ \\
    \hline
         $|V_{cb}|$                                   & $10^{-3}$    & $41.1\pm 0.5$                 & $39.5\pm0.8$      \\
         $|V_{ud}| f_\pi$                           & \text{MeV}   & $127.13 \pm 0.13$             & $126.8\pm1.4$     \\
         $|V_{us}| f_K$                             & \text{MeV}   & $35.09 \pm 0.06$              & $35.06\pm0.15$    \\
    \hline
         $\tau_{B_d}$                               & \text{ps}    & $1.519\pm 0.004$              & $1.520 \pm 0.004$ \\
         $\tau_{B_s}$                               & \text{ps}    & $1.510\pm 0.004$              & $1.505 \pm 0.004$ \\
    \hline
         $\mathcal{B}(\bar{B}^0\to D^+ K^-)$          & $10^{-3}$    & $0.326\pm 0.015$              & $0.301_{-0.031}^{+0.032}$\\
         $\mathcal{B}(\bar{B}^0\to D^{*+} K^-)$       & $10^{-3}$    & $0.327_{-0.034}^{+0.039}$      & $0.259_{-0.037}^{+0.039}$\\
         $\mathcal{B}(\bar{B}^0_s\to D^+_s \pi^-)$    & $10^{-3}$    & $4.42 \pm 0.21$               & $4.39_{-1.19}^{+1.36}$\\
         $\mathcal{B}(\bar{B}^0_s\to D^{*+}_s \pi^-)$ & $10^{-3}$    & $4.30_{-0.8}^{+0.9}$          & $2.24_{-0.50}^{+0.56}$\\
    \botrule
    \end{tabular}
    \caption{
    \label{tab:theory-inputs+results}
    Numerical inputs and results for the QCDF expressions for the branching fractions
    at leading power. We compare the results of ref.~\cite{Huber:2016xod} with our results.
    The predictions for the $\bar{B}_s^0$ branching fractions are \emph{not} time-integrated, and therefore differ
    from the measured branching fractions by a factor of $(1-y_s^2)$ \cite{DeBruyn:2012wj}.}
\end{center}
\end{table}

\section{Challenging present measurements}
\label{sec:pheno}

We compare our predictions obtained in the previous section with existing data,
and aim for a determination of $f_s/f_d$ at the LHCb experiment and the Tevatron
experiments CDF and D0. This is not trivial, for the following reasons:
\begin{enumerate}
    \item We face a circular dependence between some measurement of branching fractions and $f_s/f_d$:
    for the $\bar B_s^0$ decays in question, the values entering the world average~\cite{Tanabashi:2018oca} are at least partially using $f_s/f_d$,
    which we are aiming to extract. In \reftab{resultsBR} we make this dependence explicit, and
    introduce two independent quantities for the LHCb and CDF experiments, to highlight a potential dependence on the transverse
    momentum and hence the experiment.
    
    \item Besides correlations due to $f_s/f_d$, the measurements of the decays in question exhibit
    additional sizeable correlations.
    The largest ones are introduced because LHCb only measures ratios of branching fractions,
    albeit with very high precision.
    Furthermore, in almost all measurements the same few $D_{(s)}$ decay modes are used. Finally, we include explicitly the production fraction of neutral $B$ mesons entering absolute branching fraction measurements by the $B$ factories, since in this case isospin symmetry can be violated sizeably.
\end{enumerate}
Based on these considerations, we summarize all relevant experimental results in \reftab{exp} in the Appendix,
explicitly highlighting all cross dependence.
The individual branching fractions, their ratios, and the different production fractions presented in \reftab{resultsBR}
are produced from a series of fits to the data listed in \reftab{exp}.
In these fits, we allow for independent production fractions $f_s/f_d$ at Tevatron, LHCb (7 TeV) and LHCb (13 TeV).
The latter does not enter any measurement here and is only given for comparison.
Since only one measurement by CDF is available for the considered modes,
this effectively determines a value for $f_s/f_d$ at the Tevatron experiments,
which can then be compared to other determinations. We consider therefore the following fit scenarios:
\begin{enumerate}
    \item We first perform a fit without any input on $f_s/f_d$. The absolute scale of all $\bar B_s^0$ branching fractions is in this case provided by the measurement of $\mathcal{B}(\bar B_s^0\to D_s^+\pi^-)$ at Belle \cite{Louvot:2008sc}, which yields a very low value for $f_s/f_d$ with large uncertainties.
    \item We next include the $f_s/f_d$ value from LHCb measured at 7 TeV in semileptonic decays \cite{Aaij:2011jp,LHCb:2013lka}, which uses a method independent from the one investigated here. We make the dependence on the $D_s$ branching fraction explicit also in this case, since it is an important uncertainty and correlated with the other measurements in the fit. The dependence on $\mathcal{B}(D^-\to K^+\pi^-\pi^-)$ is not as easily included and its contribution to the uncertainty not as large as for the $D_s$ case.
\end{enumerate}
The fit results for these two scenarios are collected in \reftab{resultsBR} under ``our fits (w/o QCDF)''. Both fits describe the available data perfectly, meaning there are no obvious inconsistencies among the measurements. 
We observe significant shifts compared to the PDG fit results for $\bar B^0\to D^+ K^-$ and $\bar B^0\to D^+ \pi^-$ modes, but overall our results are well compatible with the PDG fit. More importantly, we obtain the full correlation matrix for these branching fractions, which allows to calculate their ratios with reduced uncertainties. Our improvements significantly sharpen the pattern that was apparent already in refs.~\cite{Huber:2016xod,Beneke:2000ry}: the ratios of branching fractions are well reproduced, the largest difference between measurement and prediction is 1.3$\sigma$ for $\mathcal R^{V/P}_s$. On the other hand, what was a tendency to overestimate the individual branching fractions in the past, is now a clear discrepancy: naively we observe a $4\sigma$ difference between prediction and measurement in $\bar B_s^0\to D_s
^+\pi^-$, over $5\sigma$ difference in $\bar B^0\to D^+K^-$, about $2\sigma$ in $\bar B_s^0\to D_s^{*+}\pi^-$ and $3\sigma$ in $\bar B^0\to D^{*+}K^-$. A fit to the same data as above, but expressing all branching fractions by their QCDF expressions without allowing for corrections results in $\chi^2_{\rm min}=38.7$ for 9~degrees of freedom. We see the following possibilities to resolve this discrepancy:

\begin{table*}[t]
    \centering
    \begin{tabular}{l|c cc|cc|c}
        \toprule
        source                                    & PDG               & \multicolumn{2}{c|}{our fits (w/o QCDF)}                      & \multicolumn{2}{c|}{our fit (w/ QCDF, no $f_s/f_d$)}                        & QCDF prediction\\
        scenario                                  & ---               & no $f_s/f_d$      & $(f_s/f_d)_{\rm LHCb, sl}^{\rm 7~TeV}$    & ratios only               & $\cancel{SU(3)}$                  & ---  \\
        \hline
        $\chi^2/{\rm dof}$                        & ---               & 2.5/4                             & 3.1/5                     & 4.6/6                     & 3.7/4                             & ---  \\
        \hline
        $\mathcal{B}(\bar B_s^0\to D_s^+\pi^-)$   & $3.00\pm0.23$     & $3.6\pm0.7$                       & $3.11\pm0.25$             & $3.11^{+0.21}_{-0.19}$    & $3.20^{+0.20}_{-0.26}$ ${}^*$     & $4.42 \pm 0.21$ \\
        $\mathcal{B}(\bar B^0\to D^+K^-)$         & $0.186\pm0.020$   & $0.222\pm 0.012$                  & $0.224\pm0.012$           & $0.227\pm0.012$           & $0.226\pm0.012$                   & $0.326 \pm 0.015$ \\
        $\mathcal{B}(\bar B^0\to D^+\pi^-)$       & $2.52\pm0.13$     & $2.71\pm0.12$                     & $2.73\pm0.12$             & $2.74\pm0.12$    & $2.73^{+0.12}_{-0.11}$ & ---\\
        $\mathcal{B}(\bar B_s^0\to D_s^{*+}\pi^-)$& $2.0\pm0.5$       & $2.4\pm0.7$                       & $2.1\pm0.5$               & $2.46^{+0.37}_{-0.32}$    & $2.43^{+0.39}_{-0.32}$            & $4.3^{+0.9}_{-0.8}$\\
        $\mathcal{B}(\bar B^0\to D^{*+}K^-)$      & $0.212\pm 0.015$  & $0.216\pm 0.014$                  & $0.216\pm0.014$           & $0.213^{+0.014}_{-0.013}$ & $0.213^{+0.014}_{-0.013}$         & $0.327^{+0.039}_{-0.034}$\\
        $\mathcal{B}(\bar B^0\to D^{*+}\pi^-)$    & $2.74\pm 0.13$    & $2.78\pm0.15$                     & $2.79\pm0.15$             & $2.76^{+0.15}_{-0.14}$    & $2.76^{+0.15}_{-0.14}$            & ---\\
        \hline
        $\mathcal R_{s/d}^P$                      & $16.1\pm2.1$      & $16.2\pm3.3$                      & $14.0\pm1.1$              & $13.6\pm0.6$              & $14.2^{+0.6}_{-1.1}$ ${}^*$       & $13.5^{+0.6}_{-0.5}$\\
        $\mathcal R_{s/d}^V$                      & $9.4\pm2.5$       & $11.4\pm3.6$                      & $9.6\pm2.5$               & $11.4^{+1.7}_{-1.6}$      & $11.4^{+1.7}_{-1.5}$ ${}^*$       & $13.1^{+2.3}_{-2.0}$\\
        $\mathcal R_s^{V/P}$                      & $0.66\pm0.16$     & $0.66\pm0.16$                     & $0.66\pm0.16$             & $0.81^{+0.12}_{-0.11}$    & $0.76^{+0.11}_{-0.10}$            & $0.97^{+0.20}_{-0.17}$\\
        $\mathcal R_d^{V/P}$                      & $1.14\pm 0.15$    & $0.97\pm0.08$                     & $0.97\pm0.08$             & $0.97\pm0.06$             &  $0.95\pm0.07$                    & $1.01 \pm 0.11$\\
        \hline
        $(f_s/f_d)_{\rm LHCb}^{\rm 7~TeV}$        & ---               & $0.223^{+0.056}_{-0.038}$ {}$^*$  & $0.260\pm0.019$           & $0.261^{+0.018}_{-0.016}$ & $0.252^{+0.023}_{-0.015}$ ${}^*$  & ---\\
        $(f_s/f_d)_{\rm Tev}$                     & ---               & $0.208^{+0.056}_{-0.038}$ {}$^*$  & $0.243\pm0.028$           & $0.244^{+0.026}_{-0.023}$ & $0.236^{+0.026}_{-0.022}$ ${}^*$  & ---\\
        \hline
        $\Delta_P$                                & ---               & ---                               & ---                       & $-0.164^{+0.030}_{-0.028}$ & $-0.167\pm{0.029}$               & ---\\
        $\Delta_V$                                & ---               & ---                               & ---                       & $-0.20^{+0.06}_{-0.05}$ & $-0.20^{+0.06}_{-0.05}$             & ---\\
        \botrule
    \end{tabular}
    \caption{Our fits to the available data listed in Table~\ref{tab:exp} with and without constraints from QCDF, in comparison to the PDG values \cite{Tanabashi:2018oca} and our QCDF predictions.
    The branching fractions are given in units of $10^{-3}$. Results marked with a {}$^*$ indicate that the  distribution is non-gaussian.
    The two fits on the left are not using the assumption that QCDF holds. The two fits on the right are using QCDF input to varying degree, see text. Correlations for these results are available upon request from the authors.
    }
    \label{tab:resultsBR}
\end{table*}

\begin{enumerate}
    \item One obvious option is the presence of large non-factorizable contributions of $\mathcal O(15-20\%)$ at amplitude level in each of the modes.
    This was already discussed in ref.~\cite{Huber:2016xod}, where the discrepancy has a smaller statistical significance.  When taking our new estimates in  eqs.~\eqref{eq::NLP1}~--~\eqref{eq::NLP2}, which allow already for an enhancement by a factor of 10 in the hadronic matrix elements, at face value, this scenario is clearly and significantly disfavoured at the $4.4\sigma$ level. We emphasize that we do not only see no enhancement in our calculation of next-to-leading power contributions, but instead a systematic suppression by $C_1/a_1\sim-1/3$, which renders our result particularly small. Therefore even the generic expectation of $\Lambda_\text{QCD}/m_b\sim 10\%$ seems already on the high side. We pursue this scenario nevertheless, which still allows us to extract $f_s/f_d$, albeit with increased uncertainties.
    \item We entertain also the possibility that this is an experimental issue. For that it is interesting to note that the fit to the QCDF predictions becomes excellent as soon as the measurements of the absolute branching fractions $\bar B^0\to D^{(*)+}\pi^-$ are excluded from the fit. Both values are dominated by the BaBar analysis \cite{Aubert:2006cd}, however, even the less precise CLEO data are already in conflict with the QCDF 
    predictions: the fit without these normalization modes and without input on $f_s/f_d$ yields $\mathcal B(\bar B^0\to D^+\pi^-)=(3.94^{+0.24}_{-0.20})10^{-3}$ and $\mathcal B(\bar B^0\to D^{*+}\pi^-)=(3.98^{+0.38}_{-0.37})10^{-3}$. This option would therefore imply a serious experimental problem in more than one experiment. We do not consider this option further. Nevertheless, we would like to encourage additional measurements of absolute branching fractions, which is possible with existing data from the Belle experiment and upcoming data at Belle~II. 
    \item The parametric inputs in eqs.~\eqref{eq:amp:leading-power} are in principle also potential sources of systematic shifts. However, all of them are very well known from several measurements. The quantity $|V_{cb}|$ has the largest uncertainty, but a $20\%$ shift would be in direct contradiction with all existing analyses and the global picture of the
    CKM unitarity fit.
    \item If we assume the experimental results to be correct and assume our estimates in eqs.~\eqref{eq::NLP1}-\eqref{eq::NLP2} to have the correct order of magnitude,
    only beyond the Standard Model (BSM) effects can explain the data.
    The fact that all QCDF predictions are above the corresponding measurements requires this BSM physics to interfere with the SM.
    Since the ratios are predicted well, an approximately universal factor in $b\to c\bar u d$ and $b\to c\bar u s$ transitions is preferred.
\end{enumerate}
Also a combination of the above effects is possible. In the following we discuss options 1 and 4 in detail.

\subsection{Extracting $\boldsymbol{f_s/f_d}$}
Allowing for non-factorizable contributions beyond the size indicated by the QCDF results in all decays at hand, we parametrize
\begin{align}
\label{eq::Ampnf1}
    \frac{\mathcal A(\bar B^0\to D^+K^-)}{\left.\mathcal A(\bar B^0\to D^+K^-)\right|_{\rm QCDF,LP}} &= 1+\Delta_P\,,\\
    \frac{\mathcal A(\bar B^0_s\to D^+_s\pi^-)}{\left.\mathcal A(\bar B^0_s\to D^+_s\pi^-)\right|_{\rm QCDF,LP}} &= 1+r_{SU(3)}^P\Delta_P\,,\\
    \frac{\mathcal A(\bar B^0\to D^{*+}K^-)}{\left.\mathcal A(\bar B^0\to D^{*+}K^-)\right|_{\rm QCDF,LP}} &= 1+\Delta_V\,,\\
    \frac{\mathcal A(\bar B^0_s\to D^{*+}_s\pi^-)}{\left.\mathcal A(\bar B^0_s\to D^{*+}_s\pi^-)\right|_{\rm QCDF,LP}} &= 1+r_{SU(3)}^V\Delta_V\,.
\label{eq::Ampnf4}
\end{align}
Here we use the leading-power QCDF amplitudes as in eqs.~\eqref{eq:amp:leading-power}, $\Delta_{P,V}$ parametrize non-factorizable contributions to the amplitude (including the parts estimated above) and $r_{SU(3)}^{P,V}$ parametrizes $SU(3)$ breaking beyond that in the leading amplitude. We expect $r_{SU(3)}^{P,V}\approx 1$, \emph{i.e.}, the non-factorizable parts to still scale as the decay constants, which is justified by the analytic structure of the NLP results. We choose all four parameters $\Delta_{P,V},r_{SU(3)}^{P,V}$ real without phenomenological consequences, since we only consider branching fractions here.\\

Leaving all four parameters in eqs.~\eqref{eq::Ampnf1}-\eqref{eq::Ampnf4} arbitrary is equivalent to not using the QCDF calculation at all. This reproduces the previous fit without QCDF and no input for $f_s/f_d$ in \reftab{resultsBR}.
Setting $r_{SU(3)}^{P,V}\equiv 1$, but leaving $\Delta_{P,V}$ arbitrary corresponds to the assumption that the ratios $\mathcal R_{s/d}^{P,V}$ are perfectly predicted by QCDF, while the individual branching fractions receive large non-factorizable contributions. This mimics one of the assumptions employed in ref.~\cite{Fleischer:2010ca,Aaij:2013qqa} and in particular allows for a comparison with the result in ref.~\cite{Aaij:2013qqa}. We list the results of this scenario in \reftab{resultsBR} under ``ratios only''. However, given the significant reduction of the parametric theory uncertainty in this work, the assumption of a negligible uncertainty from $SU(3)$ breaking in the non-factorizable part does not seem appropriate anymore. While we do not use the quantitative results for the NLP contributions here, we still make observations from their analytical structure:
\begin{enumerate}
    \item In general there is no justification to identify $\Delta_P$ and $\Delta_V$.
    \item The deviation of $r_{SU(3)}^{P,V}$ from unity is expected to be small: the sizable breaking from the light-meson decay constants is identical to that of the leading amplitude, and the heavy-to-heavy matrix elements for a single soft gluon at NLP exhibit a similar structure to that of the form factors, for which the results in ref.~\cite{Bordone:2019guc} show explicitly that the breaking is small, in accordance with theoretical expectations from the HQE \cite{Jenkins:1992qv,Boyd:1995pq,Kobach:2019kfb}. We therefore consider $r_{SU(3)}^{P,V}\in [0.9,1.1]$ to be a conservative estimate of this breaking, leaving the two parameters independent. 
\end{enumerate}
We consequently perform another fit, leaving $\Delta_{P,V}$ independent and arbitrary while constraining the $SU(3)$ breaking beyond that in the leading amplitude to be below $10\%$. In this way we account for the possibility of relevant additional $SU(3)$ breaking, while still exploiting the information from QCDF. The results of this fit are reported again in \reftab{resultsBR}. 

Since this scenario interpolates in a way between the fit without QCDF and without $f_s/f_d$ input and the one without $SU(3)$ breaking, it is not surprising that also the results for $f_s/f_d$ lie between the corresponding values. As mentioned before, we observe large non-factorizable contributions of $15-20\%$, independently of the allowed magnitude of $SU(3)$ breaking. Although the minimal $\chi^2$ decreases by $0.9$ when allowing for $SU(3)$ breaking, such a fit has two fewer degrees of freedom, so there is no indication for sizable $SU(3)$ breaking in the data. We consider this scenario nevertheless preferable over the one without breaking, since the two parameters $r_{SU(3)}^{P,V}$ account for the related uncertainty and render our results for $f_s/f_d$ conservative. While this is not the only way to estimate this uncertainty, the sizable shift between the central values and uncertainties in the  scenarios with and without breaking, despite the relatively small breaking beyond that of the leading amplitude of $\leq10\%$, indicates that this source of uncertainty must be taken into account for a meaningful extraction of $f_s/f_d$.\\

Our results are compatible with previous extractions of $(f_s/f_d)_{\rm LHCb}^{\rm 7\,TeV}$ from $\mathcal R_{s/d}$ \cite{Aaij:2011hi,Bailey:2012rr,LHCb:2013lka,Aaij:2013qqa,Monahan:2017uby}, but have significantly reduced uncertainties, despite allowing for larger $SU(3)$ breaking in the non-factorizable part. Both our values are also in excellent agreement with the value from semileptonic decays. The uncertainties are comparable, and can be reduced with more data in the future. However, the problem of the large apparent corrections of the QCDF results remains. Furthermore, while we consider our estimate for the range of $r_{SU(3)}^{P,V}$ to be conservative, it is clear from the data in \reftab{resultsBR} that the result for $f_s/f_d$ sensitively depends on our assumption. An effort should therefore be made to even further improve the understanding of these modes, in order to either understand the source for these large non-factorizable contributions, or to establish the presence of BSM physics in these modes, as discussed in the next subsection.

Finally, let us comment on the value for $(f_s/f_d)_{\rm Tev}$ extracted in \reftab{resultsBR}: the value quoted in ref.~\cite{Amhis:2019ckw} reads $(f_s/f_d)_{\rm Tev}=0.334\pm0.040$, which differs by $\sim2\sigma$ from the value obtained in both fits using QCDF in \reftab{resultsBR}. It is worth emphasizing that our value is independent from the values entering that average, and more precise. On the other hand it relies on LHCb-data and is hence not a pure CDF measurement. It is not clear from the information provided in ref.~\cite{Amhis:2019ckw} how the average is obtained. Updating the external inputs for the analyses \cite{Affolder:1999iq,Aaltonen:2008zd} and performing a correlated average, we obtain 
\begin{equation}
    (f_s/f_d)_{\rm Tev,sl}=0.263\pm0.031\,,
\end{equation}
which is also more precise than the average quoted in ref.~\cite{Amhis:2019ckw} and perfectly compatible with both the LHCb results (at different transverse momentum) and our result from non-leptonic decays.

We therefore conclude that while we do not see whence the required large non-factorizable contributions could originate, assuming their presence yields a consistent picture for all available data and an improved determination of $f_s/f_d$, which can be improved further in the future.

\subsection{New physics in $\boldsymbol{b\to c\bar u s(d)}$ transitions}

We now explore the possibility that the discrepancy between data and the SM calculation is caused by BSM physics. This is in part motivated by the observation that our fits including the QCDF constraints allow for $\Delta_P=\Delta_V$, although this would not be expected for NLP contributions. It would, however, be expected for BSM contributions to the Wilson coefficients $C_{1,2}^{q_2}$.
The ratios $\mathcal R_{d,s}^{V/P}$ therefore provide tests for our BSM hypothesis.

We do not aim at a detailed BSM analysis, only at establishing whether this option is already ruled out by existing data.\\

The observation that the new contributions should similarly reduce branching fractions with a pseudoscalar meson $D_q$ and a vector meson $D_q^*$ has rather strong consequences, implying specifically a
minimal-flavour-violation-like scenario where the BSM contributions scale with the CKM factors of the SM ones. 
Scalar BSM operators generate $\Delta_P \sim 1/(m_b-m_c)$, and $\Delta_V\sim 1/(m_b+m_c)$. This asymmetry does not exclude solutions with scalar operators, but requires a careful treatment of the various BSM contributions. For simplicity we content ourselves with the obvious option that BSM physics only modifies the SM Wilson coefficients, which yields $\Delta_P= \Delta_V$.
 More precise data will provide a test of this assumption. Using this assumption, we obtain a universal shift in $a_1$ for
each class of transitions $b\to c\bar u q_2$, \emph{i.e.}, we have in general two contributions $\Delta a_1^{q_2}$. Again for simplicity, we use the observation that the data allow for
$\Delta a_1^d=\Delta a_1^s\equiv \Delta a_1$ to simplify our analysis further. Clearly this assumption can be easily tested in the ratios $\mathcal R_{s/d}^{P,V}$. This scenario corresponds
to the one above in the limit $\Delta_P=\Delta_V\equiv \Delta \tilde a_1=\Delta a_1/a_1$, since $a_1$ to very good approximation is universal. 

As suggested by the previous fits, our BSM scenario works very well: we obtain $\chi^2_{\rm min}=4.9$ for 7 degrees of freedom. We also obtain $(f_s/f_d)_{\rm LHCb}^{\rm 7\, TeV}=0.263^{+0.018}_{-0.016}$, which has the same uncertainty as in the ``ratios only'' scenario; however, clearly this value depends sensitively on our assumption of $\Delta a_1^d=\Delta a_1^s$. Most importantly, we obtain
\begin{equation}
    \Delta \tilde a_1 = -0.17\pm0.03,\quad\text{or}\quad \Delta  a_1 = -0.18\pm0.03\,,
\end{equation}
ignoring potential imaginary parts for now using the same justification as given above.

We check if our BSM physics hypothesis is already excluded by other observables. In the following we list these observables and discuss the impact of their measurements, see \reftab{NP}:
\begin{itemize}
    \item $\Gamma_{q}$: since we modify the leading contribution to the total decay rates of the $\bar B_q$ mesons, we expect a strong constraint from $\Gamma_q$, $q=d,s$. However, each individual total decay rate is 
    not as precisely predicted as their ratio.
    \item $\tau_{B_s}/\tau_{B_d}$: this observable is both predicted and measured to very high precision. The main contributions to the individual lifetimes cancel in the ratio, therefore the dominant contribution in our scenario is via dimension-6 operators in the OPE to $\tau_{B_d}$. We calculate the BSM shift at leading order, using the results from ref.~\cite{Neubert:1996we} with updated bag factors \cite{Kirk:2017juj}.
    \item $a_d^{fs}$: the flavour-specific CP-asymmetry in the $B_d-\bar B_d$ system receives also a leading contribution from $b\to c\bar ud$ operators, which is linear in the corresponding coefficient \cite{Bobeth:2014rda}. This constraint is therefore complementary to the others.
\end{itemize}
We do not consider further observables that are not as cleanly predicted as the ones above. We note, however, that other $B$ meson decays like $\bar B^0\to D^{(*)+}\pi^-$ and $\bar B_{s}\to D_{s}^{(*)+}K^-$ (and their analogues with $\rho,K^*$) are likewise affected by our BSM physics hypothesis. We emphasize that \emph{all} corresponding measurements \cite{Tanabashi:2018oca} are lower than their (updated) SM predictions~\cite{Huber:2016xod}, thereby strengthening the case for
the BSM hypothesis.

\begin{table}[th]
    \centering
    \begin{tabular}{l c c l}\hline\hline
         observable                 & measurement       & SM prediction     & ref.\\\hline
         $\Gamma_d/10^{-13}$~GeV    & $4.333\pm0.011$   & $3.6\pm0.8$       & \cite{Bobeth:2014rda,Krinner:2013cja}\\
         $\tau_{B_s}/\tau_{B_d}$    & $0.994\pm0.004$   & $1.0006\pm0.0020$ & \cite{Amhis:2019ckw,Kirk:2017juj}\\
         $a^{fs}_d/10^{-4}$         & $-21\pm17$        & $-4.73\pm0.42$    & \cite{Amhis:2019ckw,Lenz:2019lvd}\\
        \hline\hline 
    \end{tabular}
    \caption{Measurements and predictions for further observables that constrain BSM physics in $b\to c\bar u (d,s)$.}
    \label{tab:NP}
\end{table}

We find that surprisingly none of the inclusive observables in \reftab{NP} excludes the possibility of having a shift of $-15\%$ to $-20\%$ in $a_1$ from BSM physics,
in line with refs.~\cite{Bobeth:2014rda,Lenz:2019lvd}. We conclude that BSM physics is a viable possibility that could explain the observed puzzle. Despite requiring a large contribution of $\sim -17\%$ of a CKM-leading tree-amplitude, available constraints from the above inclusive observables do not exclude such a scenario, which therefore has to be taken seriously. A detailed BSM analysis of these constraints is, however, beyond the scope of this article and left for future work.

\subsection{Prospects}

Our results reduce both main uncertainties in the predictions for $\bar{B}_{(s)}^0 \to D_{(s)}^+ \lbrace \pi^-, K^-\rbrace$ decays, thereby enabling future precision analyses of these modes. Measurements of absolute branching fractions can and should be carried out at Belle~II, if possible also for the $\bar B_s$ decays modes. High-precision measurements of ratios of branching fractions can be obtained by the LHC experiments. With further reduced uncertainties, it would also be helpful to have smaller uncertainties on the $D_{(s)}^+$ branching fractions.

Experimentally, it might be advantageous to normalize to a different mode, like $\bar B^0\to D^{(*)+}\pi^-$, since it has a larger branching fraction and systematic uncertainties due to the pions cancel. As an example, the prediction for the ratio $\mathcal{B}(\bar B_s^0\to D_s^{+}\pi^-)/\mathcal{B}(\bar B^0\to D^{+}\pi^-)$ involves a different ratio of form factors than the ones already provided. Its value obtained from ref.~\cite{Bordone:2019guc},
\begin{equation}
  \left\vert  \frac{F_0^{\bar {B}_s\to D_s}(m_\pi^2)}{F_0^{\bar {B}\to D}(m_\pi^2)} \right\vert = 1.005\pm 0.021\,,
\end{equation}
is rather precise. We caution that these modes suffer from significantly larger uncertainties due to presently unquantifiable $\mathcal O(\Lambda_\text{QCD}/m_b)$ corrections. Therefore, the theoretically cleanest ratios for the purpose of determining $f_s/f_d$ are $\mathcal{R}_{s/d}^{P/V}$, discussed in the previous sections.  
 
As a workaround, we suggest a two-staged approach:
the measurement of $\mathcal{B}(\bar B\to X)/\mathcal{B}(\bar B^0 \to D^+ K^-)$ --- which, depending on the choice of $X$, can be determined at a precision of a few percent and potentially further improved at Belle~II or LHCb --- can convert the more easily measurable ratio $\mathcal{B}(\bar{B}_s^0 \to D_s^+ \pi^-) / \mathcal{B}(\bar B\to X)$ into one of $\mathcal{R}_{s/d}$, \emph{i.e.},
\begin{align}
    \mathcal R_{s/d}^P &= \frac{\mathcal B(\bar B_s\to D_s^+\pi^-)}{\mathcal B(\bar B\to X)}\frac{\mathcal B(\bar B\to X)}{\mathcal B(\bar B^0\to D^+K^-)}\\
    &=\frac{\mathcal B(\bar B_s\to D_s^+\pi^-)}{\mathcal B(\bar B\to X)} R_{d}^{X}\,.
\end{align}
For instance, with $X=D^+\pi^-$ the required experimental ratio would already be available at the $3-4\%$ level:
\begin{equation}
    R_{d}^{\pi K} \equiv \frac{\mathcal B(\bar B^0\to D^+\pi^-)}{\mathcal B(\bar B^0\to D^+K^-)} \stackrel{\rm exp}{=} 12.17^{+0.42}_{-0.37}.
\end{equation}
The situation is similar for $X= D^0\pi^-$; the determination of the ratio $\mathcal R_{ud}^{D^0\pi^-}=\mathcal B(B^-\to D^0\pi^-)/\mathcal B(\bar B^0\to D^+ K^-)$ requires additionally the determination of the production fraction of charged and neutral $B$ meson pairs at the $B$ factories, or the ratio $f_u/f_d$ at LHCb (which is however expected to be close to unity). We are looking forward to see these prospects realised in future analyses by the ATLAS, Belle~II, CMS and LHCb experiments.

\section{Discussion and Conclusion}
\label{sec:conclusion}

In this work we revisit, extend and update the theoretical predictions for the branching fractions of the decays $\bar{B}^0\to D^{(*)+} K^-$ and $\bar B_s^0\to D^{(*)+}_s\pi^-$.
We obtain our predictions in the framework of QCD factorization (QCDF), using next-to-next-to leading order results for the effective Wilson coefficients and including next-to-leading power (NLP) corrections for the first time.  Moreover, we employ updated values of the form factors with reduced theoretical uncertainties compared to previous estimates.
Beyond the prediction of the branching fractions, we provide updated results for the fragmentation fraction $f_s/f_d$ in various scenarios. To obtain a reliable error estimate for the fragmentation fraction, we consider it mandatory to account for potential $SU(3)$ breaking in the non-factorizable corrections. Our results agree with the values extracted by LHCb and CDF from semileptonic decays.

Our comparison of the various experimental measurements of these modes and our theoretical predictions shows a clear and very significant discrepancy at the level of $4.4\sigma$. This high level is due to drastically reduced parametric uncertainties.
We identify the following possible four causes of the discrepancy, none of which is fully satisfactory on its own:

\begin{enumerate}
    \item The current measurements of the absolute branching fractions for the modes considered could have a systematic bias in form of a downward shift.\\
    This is unlikely to be the case, since the modes we discuss have large branching fractions, and have only charged particles in the final state, rendering them experimentally well accessible.
    A systematic bias of the order of $\sim -30\%$ would lead to questioning the validity of all measurements of $B$ meson branching fractions.
    
    \item The theoretical results within the framework of QCDF could miss a large contribution of $\sim -20\%$ at the amplitude level in color-allowed tree topologies.\\
    This is unlikely to be case, since QCDF predictions for color-allowed tree decays into light mesons are in reasonable agreement with the measurements. If such a shift in heavy-light final states is necessary it is hard to understand why a similarly sized shift in light-light final states is absent.
    
    \item The hadronic matrix elements for the non-local soft gluon terms could be underestimated, and cause the observed shift.\\
    To increase the $\bar{B}\to D^{(*)}$ matrix elements to the size needed for agreement with the measurements, we would need an enhancement by a factor of $\sim 50$.
    This is also quite unlikely to be the case, since we have already enlarged the possible range of our estimates by multiplying our nominal results by a factor of 10 to obtain conservative estimates.

    \item 
    The discrepancy could be caused by contributions from physics beyond the SM.\\
    Also this option seems unlikely, given that 
    the partonic transition is generated by a tree-level $W$ exchange, and it would be surprising if a correction of $\sim -20\%$  had eluded attention so far. 
    On the other hand, the same pattern exists for $b\to c\bar u(d/s)$ transitions that are not quantitatively discussed in this work. The interpretation is not excluded either by a number of inclusive observables like the flavor-specific CP asymmetry $a_{fs}^d$, the total width of the $B$ meson $\Gamma_d$, or the ratio of lifetimes $\tau_s/\tau_d$.
\end{enumerate}
Given the significance of this puzzle and the potential impact of any of the potential explanations discussed above, each of them should be studied in more detail.
Doing so exhaustively requires dedicated efforts on both the experimental and the theoretical side.
We look forward to further study these options in detail in a forthcoming publication.

\acknowledgments

We thank Greg Landsberg for triggering our interest in this prediction, and both Greg Landsberg and Mick Mulder for communication on the measurements at the CMS and LHCb experiments, respectively. We also thank Martin Beneke and Christoph Bobeth for useful discussions. DvD is grateful to K.~Keri Vos for helpful discussions. TH would like to thank Xin-Qiang Li for useful correspondence. MB thanks Admir Greljo for useful correspondence.

The work of MB and TH is supported by Deutsche Forschungsgemeinschaft (DFG, German Research Foundation) under grant 396021762 - TRR 257 ``Particle Physics Phenomenology after the Higgs Discovery''.
The work of NG and DvD is supported by the DFG within the Emmy Noether Programme under grant DY130/1-1 and
the DFG Collaborative Research Center 110 ``Symmetries and the Emergence of Structure in QCD''.
The work of MJ is supported by the Italian Ministry of Research (MIUR) under grant PRIN 20172LNEEZ.
This research was supported by the Cluster of Excellence ``ORIGINS'' and the Munich Institute for Astro- and Particle Physics (MIAPP) which are funded by the Deutsche Forschungsgemeinschaft (DFG, German Research Foundation) under Germany's Excellence Strategy --- EXC-2094 --- 390783311.

\appendix

\section{Light-cone sum rule for the soft-gluon matrix element}
\label{app:soft-gluon}

The next-to-leading power term in the factorization formula for $\bar{B}_q\to D_q^{(*)+} L^-$ involves an exchange of a soft-gluon between
the heavy-to-heavy transition and the light-meson.
This exchange can be described in terms of
the hadronic matrix elements of the operator~\cite{Beneke:2000ry,Beneke:2000pw}
\begin{equation}
    \mathcal{O} \equiv -2\int_0^{1/\Lambda_\text{QCD}} ds\, \bar{c}(0) \gamma^\mu (1 - \gamma_5) \tilde{G}_{\mu\nu}(-s n) n^\nu b(0)\,,
\end{equation}
where $n$ is a light-cone vector and $q_\mu \simeq E_L n_\mu$ in the $B$-meson rest frame.
Note that we absorb a factor of $-1$ into the definition of the operator above.
The explicit factor of two accounts for the difference between the definition of $\tilde{G}$ in refs.~\cite{Beneke:2000ry,Beneke:2000pw}
and our definition
\begin{equation}
    \tilde{G}_{\mu\nu} \equiv \frac{1}{2} \varepsilon_{\mu\nu \alpha\beta} G^{\alpha\beta}\,,
\end{equation}
which agrees with the one in refs.~\cite{Khodjamirian:2010vf,Braun:2017liq}.
Matrix elements of the operator $\mathcal{O}$ can be estimated in a similar fashion as done
for the matrix elements at subleading power for the non-local contributions in exclusive
$b\to s\ell^+\ell^-$ decays~\cite{Khodjamirian:2010vf}. We closely follow the procedure to estimate
the latter.\\

Using the translation operator, we can relate the gluon field on the light cone to the gluon field at the origin:
\begin{equation}
    G_{\mu\nu}(-s n) = \exp\left[-i (-s n)\cdot (iD)\right] G_{\mu\nu}(0)\,.
\end{equation}
This step allows us to express $\mathcal{O}$ as follows:
\begin{align}
    \mathcal{O}
        & = -2\int_0^\infty d\omega_2 \, \int_0^{1/\Lambda_\text{QCD}} ds\, e^{i s \omega_2}\\
    \nonumber
        & \times \bar{c}(0) \gamma^\mu (1 - \gamma_5) \delta\left[\omega_2 - (n\cdot i D)\right]\tilde{G}_{\mu\nu}(0) n^\nu b(0)\,.
\end{align}
To leading-power in $\Lambda_\text{QCD}/E_L$ the effective integration range for $s$ is $[0, \infty)$.
This leads us to the momentum space representation
\begin{align}
    \mathcal{O}
        & = -2 i\int_0^\infty \frac{d\omega_2}{\omega_2} \\
    \nonumber
        & \times \bar{c}(0) \gamma^\mu (1 - \gamma_5) \delta\left[\omega_2 - (n\cdot i D)\right]\tilde{G}_{\mu\nu}(0) n^\nu b(0)\\
        & \equiv -2i \int_0^\infty \frac{d\omega_2}{\omega_2}\tilde{\mathcal{O}}(\omega_2) \equiv -2i \tilde{\mathcal{O}}\,,
\end{align}
where $\omega_2$ is the light-cone component of the gluon momentum.\\

We proceed to estimate the matrix elements 
\begin{align}
    \bra{D^{(*)+}(k)} \tilde{\mathcal{O}} \ket{\bar B^0(q + k)}\,.
    \label{eq:non-localME}
\end{align}
For this task we use light-cone sum rules with an on-shell $B$-meson. The relevant three-particle $B$-meson
LCDAs are defined in the heavy-quark limit and therefore only represent
the result of the matrix element of $\tilde{\mathcal{O}}$ to leading power in $\Lambda_\text{QCD}/m_b$.
We define the correlation functions
\begin{align}
    \label{eq:LCSRcorr}
	& \Pi_{D^{(*)}}(q, k)\\
    \nonumber
	& = i \, \int d^4x \, e^{i k x} \bra{0} \mathcal{T} \left\lbrace J_\text{int}^{D^{(*)}}(x), \tilde{\mathcal{O}} \right\rbrace \ket{\bar B^0(q+k)}\,, 
\end{align}
where $J^{D^{(*)}}_\text{int}(x) \equiv \bar{d}(x) \Gamma_{D^{(*)}} c(x)$ denotes a current suitable to interpolate the $D$ or $D^*$. We use
\begin{align}
    \Gamma_{D}   & = i \gamma_5\,, &
    \Gamma_{D^*} & = \slashed{q} + \slashed{k}\,.
\end{align}

Inserting a complete set of states between the two currents, we obtain the hadronic representation of the correlation functions. For the pseudoscalar meson $D = D(k)$ we obtain:
\begin{equation}
\begin{aligned}
	\Pi_D^\text{had}(q, k)
	\label{eq:disprel:D}
	    & = \frac{f_D m_D^2}{m_c}
	    \frac{
	        \bra{D^+}       \tilde{\mathcal{O}}  \ket{\bar B^0(q + k)}
	    }
	    {
	        m_D^2-k^2
	    } \\
	    & + \frac{1}{\pi}
	    \int_{s_h}^\infty d s\frac{\tilde{\rho}_D(s,q^2)}{s-k^2}
	    \,.
\end{aligned}
\end{equation}
For a longitudinal vector meson $D^{*} \equiv D^{*}(k, \lambda=0)$ we obtain:
\begin{equation}
\begin{aligned}
	\Pi_{D^*}^\text{had}(q, k)
	\label{eq:disprel:Dst}
	    & = 
	    \frac{ f_{D^*} \sqrt{\lambda_\text{kin}}}{2}
	    \frac{
	        \bra{D^{*+}}      \tilde{\mathcal{O}} \ket{\bar B^0(q + k)}
	    }
	    {
	        m_{D^*}^2-k^2
	    } \\
	    & + \frac{1}{\pi}
	    \int_{s_h}^\infty d s\frac{\tilde{\rho}_{D^*}(s,q^2)}{s-k^2}
	    \,,
\end{aligned}
\end{equation}
where $\lambda_\text{kin} \equiv \lambda(m_B^2, m_{D^*}^2, q^2)$ is the K\"all\'en function.
For both cases above $\tilde{\rho}_{D^{(*)}}$ denotes the spectral density of the respective exited and continuum states.
\\

At $q^2 \simeq 0$ the correlators \refeq{LCSRcorr} are accessible in a light-cone operator product expansion (OPE) if $k^2 < 0$, and $|k^2| \gg \Lambda_\text{QCD}^2$.
For these kinematics, the hard and soft contributions factorize.
The hard contributions are calculated to leading order in $\alpha_s$, while the soft contributions are expressed
in terms of the $B$-mesons LCDAs.
For both the $D$ and the $D^{*}$ meson, the OPE result can be conveniently expressed as:
\begin{multline}
    \raisetag{5em}
    \label{eq:OPEresult}
	\Pi_{D^{(*)}}^{\text{OPE}}(q, k)
	    = 
	    \int d\omega_2 \, \int d^4x \, \int d^4 p' \, e^{i (k - p') \cdot x}
	    \\
	        \left[
	            \Gamma_{D^{(*)}} \, \frac{\slashed{p}' + m_c}{m_c^2 - p'^2} \, \gamma^\mu (1-\gamma_5)   
	        \right]_{ab}
	   \\
	        \bra{0} \bar{d}^a(x) \delta\left[\omega_2 - i n \cdot D\right] 
	        \tilde{G}_{\mu\nu}(0) n^\nu h_v^b(0) \ket{\bar{B}^0}\,.
\end{multline}
The sum rules are then obtained by matching the OPE results (\ref{eq:OPEresult}) onto the hadronic representations (\ref{eq:disprel:D})-(\ref{eq:disprel:Dst}).
To remove continuum contributions, the assumption of semi-global quark-hadron duality approximation is used,
as discussed in ref.~\cite{Gubernari:2018wyi}.
We Borel transform the result to suppress the tail of the OPE calculations and the continuum contributions,
thereby reducing the numerical impact of violation of quark-hadron duality.
For our results we use the twist classification and modelling of the LCDAs as in ref.~\cite{Braun:2017liq}, and
truncate the twist expansion at the twist-four level.\\

Using the same inputs as in ref.~\cite{Gubernari:2018wyi}, our numerical results for $q^2 = 0$ are
\begin{align}
    \bra{D^+(k)} {\mathcal{O}}  \ket{\bar{B}^0(q + k)}
        & = i\left[0.13, 1.3\right]\,{\rm GeV}^2\,,\\
    \bra{D^{*+}(k, \eps(\lambda=0))}\mathcal{O} \ket{\bar{B}^0(q + k)}
        & = i \left[0.078, 0.78\right] \,{\rm GeV}^2\,.
\end{align}
The lower bounds in the ranges above are obtained using the nominal values of our input parameters.
To have a conservative estimate of these contributions, we compute the upper bounds by increasing the values of $\lambda_E^2$ and $\lambda_H^2$ by one order of magnitude. These two parameters enter approximately linearly in the normalization of the $B$-meson LCDAs. An increase in their central values implies an increase in our prediction of the matrix elements (\ref{eq:non-localME}).
Normalizing these non-local  matrix elements with the corresponding local matrix elements, we obtain
\begin{align}
    \frac{
        \bra{D^+(k)} \mathcal{O} \ket{\bar B^0(q + k)  }
    }{
        i(m_B^2 - m_D^2) F_0^{\bar B\to D}(0) 
    }
    & = [0.76, 7.6]\% \,,\\
    \frac{
        \bra{D^{*+}(k, \lambda = 0)} \mathcal{O} \ket{\bar B^0(q + k)}
    }{
        i(m_B^2 - m_{D^*}^2) A_0^{\bar B\to D^*}(0) 
    }
    & = [0.46,4.6]\%\,.
\end{align}
Note that here $A_0$ is defined as in ref.~\cite{Beneke:2000ry}, which differs in the phase convention from ref.~\cite{Gubernari:2018wyi}.\\

\section{Experimental inputs}
\label{app:experimentalinputs}
We update the experimental analyses to account for new results regarding, \emph{e.g.}, $D_{(s)}^-$ branching fractions, lifetimes, and production fractions, which is only partially done in the averages available. 
In \reftab{exp} we summarize the pertinent measurements; we use the information in the articles to extract the quantities listed there, which allow to explicitly account for all correlations due to $f_s/f_d$ as well as the charm branching fractions. A few comments are in order:
\begin{itemize}
    \item The branching fraction from ref.~\cite{Louvot:2008sc} depends on the production of $B_s$ in $\Upsilon(5S)$ decays. This is the only measurement where this quantity enters, so we do not make this dependence explicit.
    \item 
    The fragmentation fraction $f_s/f_d$ potentially depends on the experiment, specifically via the transverse momentum $p_T$, as observed for $f_s/f_u$ \cite{Aaij:2019eej}, $\Lambda_b$ production \cite{Amhis:2019ckw} and indicated by another LHCb measurement~\cite{Aaij:2013qqa}. On the other hand, the result from LEP seems to indicate a milder dependence on $p_T$ than found in ref.~\cite{Aaij:2013qqa}, as discussed in ref.~\cite{Amhis:2019ckw}. 
    The treatment of this dependence differs in the literature.
    The most conservative approach is to use the values extracted at the Tevatron and LHCb experiments independently, the averages of which differ sizeably.  We present the experimental results in a way that allows for different treatments, specified in the corresponding paragraphs.
    \item A third measurement of $\bar{B}\to D^-K^+$ listed in ref.~\cite{Tanabashi:2018oca} is declared superseded by LHCb in their later article. 
    \item The production fraction $f_{00}$ of $B/\bar B$ pairs in $\Upsilon(4S)$ decays is assumed to be $1/2$ in most measurements (and for all results in ref.~\cite{Tanabashi:2018oca}), but can differ sizeably from that value, see ref.~\cite{Jung:2015yma} for a recent discussion.
\end{itemize}

\begin{table*}[t]
    \centering
    \renewcommand{\arraystretch}{1.5}
    \begin{tabular}{l c c r}\hline\hline 
    measurement & value & source    & reference(s)\\\hline
        $\mathcal{B}(B_s^0\to D_s^-\pi^+)$ & $(3.6\pm0.5\pm0.5)\,10^{-3}$ & Belle & \cite{Louvot:2008sc,Tanabashi:2018oca}\\
        $\frac{f_s}{f_d}\dfrac{\mathcal{B}(B_s^0\to D_s^-(\to \phi(\to K^+K^-)\pi^-) \pi^+)}{\mathcal{B}(B^0\to D^-(\to K^+\pi^-\pi^-)\pi^+)}$ & $(6.7\pm0.5)\%$ & CDF & \cite{Abulencia:2006aa}$^*$\\
        $\frac{f_s}{f_d}\dfrac{\mathcal{B}(B_s^0\to D_s^-(\to K^+K^-\pi^-) \pi^+)}{\mathcal{B}(B^0\to D^-(\to K^+\pi^-\pi^-)\pi^+)}$ & $0.174\pm 0.007$ & LHCb & \cite{Aaij:2012zz}\\
        $\frac{f_s}{f_d}\dfrac{\mathcal{B}(B_s^0\to D_s^-(\to K^+K^-\pi^-) \pi^+)}{\mathcal{B}(B^0\to D^-(\to K^+\pi^-\pi^-)K^+)}$ & $2.08\pm 0.08$ & LHCb & \cite{Aaij:2013qqa}$^\dagger$\\
        $\dfrac{\mathcal{B}(B^0\to D^-K^+)}{\mathcal{B}(B^0\to D^-\pi^+)}$ & $(8.22\pm0.28)\%$ & LHCb & \cite{Aaij:2013qqa}$^\dagger$\\
        $\dfrac{\mathcal{B}(B^0\to D^-K^+)}{\mathcal{B}(B^0\to D^-\pi^+)}$ & $(6.8\pm1.7)\%$ & Belle & \cite{Abe:2001waa}\\
        $f_{00}\mathcal{B}(B^0\to D^-(\to K^+\pi^-\pi^-)\pi^+)$ & $(1.21\pm 0.05)\,10^{-4}$ & BaBar/CLEO & \cite{Aubert:2006cd,Ahmed:2002vm}\\
        $\mathcal{B}(B^0\to D^-(\to K^+\pi^-\pi^-)\pi^+)$ & $(2.88\pm 0.29)\,10^{-4}$ & BaBar & \cite{Aubert:2006jc}$^\S$\\
        $\dfrac{\mathcal{B}(B_s^0\to D_s^{*-}\pi^+)}{\mathcal{B}(B_s^0\to D_s^{-}\pi^+)}$ & $0.66\pm 0.16$ & Belle & \cite{Louvot:2010rd}\\
        $\dfrac{\mathcal{B}(B^0\to D^{*-}K^+)}{\mathcal{B}(B^0\to D^{*-}\pi^+)}$ & $(7.75\pm0.30)\%$ & LHCb/BaBar/Belle & \cite{Aubert:2005yt,Abe:2001waa,Aaij:2013xca}\\
        $f_{00} \mathcal{B}(B^0\to D^{*-}\pi^+)$ & $(2.72\pm0.14)\,10^{-3}$ & BaBar/CLEO & \cite{Aubert:2006cd,Brandenburg:1997zs}\\
        $\dfrac{\mathcal{B}(B^0\to D^{*-}\pi^+)}{\mathcal{B}(B^0\to D^-\pi^+)}$ & $0.99\pm0.14$ & BaBar & \cite{Aubert:2006jc}\\
        \hline
        $\mathcal{B}(D_s^-\to\phi(\to K^+K^-)\pi^-)$ & $(2.27\pm0.08)\%$      & PDG average & \cite{Tanabashi:2018oca}\\
        $\mathcal{B}(D_s^-\to K^+K^-\pi^-)$ & $(5.45\pm0.17)\%$ & PDG average & \cite{Tanabashi:2018oca}\\
        $\mathcal{B}(D^-\to K^+\pi^-\pi^-)$ & $(9.38\pm0.16)\%$ & PDG average & \cite{Tanabashi:2018oca}\\
        \hline 
        $\mathcal{B}(D_s^-\to K^+K^-\pi^-)(f_s/f_d)_{\rm LHCb, sl}^{7{\rm TeV}}$ & $0.0144\pm 0.0010$ & LHCb                       & \cite{Aaij:2011jp,LHCb:2013lka}\\
        $\mathcal{B}(D_s^-\to K^+K^-\pi^-)(f_s/f_d)_{\rm LHCb, sl}^{13{\rm TeV}}$ & $0.0133\pm 0.0005$ & LHCb                       & \cite{Aaij:2019pqz}\\
        $(f_s/f_d)_{\rm Tev}$ & $0.334\pm 0.040$                        & HFLAV average              & \cite{Amhis:2019ckw}\\
        $f_{00}$ & $0.488\pm0.010$                                      & pheno comb. of BaBar/Belle & \cite{Hastings:2002ff,Aubert:2005bq,Jung:2015yma}\\
        \hline\hline
    \end{tabular}
    \caption{Relevant experimental measurements entering our determination of branching fractions and ratios for the considered modes. A ${}^*$ indicates that here $f_s/f_d$ corresponds to the value at Tevatron, see text. The measurements marked by a ${}^\dagger$ from ref.~\cite{Aaij:2013qqa} have a correlation coefficient of $-56\%$. The reference marked with $^\S$ uses both $D^-\to K^+\pi^-\pi^-$ and $D^-\to K_S\pi^-$ decays, however, the former is dominating.}
    \label{tab:exp}
\end{table*}

\bibliography{references.bib}

\end{document}